\documentclass[10pt]{article} 
\usepackage{amssymb}

\font\tenscr=rsfs10 scaled1100
\font\sevenscr=rsfs7 
\font\fivescr=rsfs5 
\skewchar\tenscr='177 \skewchar\sevenscr='177 \skewchar\fivescr='177
\newfam\scrfam
\textfont\scrfam=\tenscr
\scriptfont\scrfam=\sevenscr
\scriptscriptfont\scrfam=\fivescr

\def\scri{{\fam\scrfam I}}
\def\N{\mathcal{N}}
\def\Z{\mathcal{Z}}
\def\S{\mathcal{S}}
\def\O{\mbox{O}}

\newtheorem{proposition}{Proposition}

\begin{document}

\title{Logarithmic Newman-Penrose constants for arbitrary polyhomogeneous
 spacetimes}

\author{Juan Antonio Valiente Kroon\thanks{%
E-mail address: j.a.valiente@qmw.ac.uk} \\
School of Mathematical Sciences, \\
Queen Mary \& Westfield College, \\
Mile End Road, London E1 4NS.}

\maketitle

\begin{abstract}
A discussion of how to calculate asymptotic expansions for polyhomogeneous 
spacetimes using the Newman-Penrose formalism is made. The existence of 
logarithmic Newman-Penrose constants for a general 
polyhomogeneous spacetime (i.e. a polyhomogeneous spacetime such that 
$\Psi_0=\O(r^{-3}\ln ^{N_3})$) is addressed. It is found that these constants 
exist for the generic case. 

\end{abstract}


\section{Introduction}

The study of the gravitational field of isolated sources close to null 
infinity $\scri$ has been done by solving the \emph{asymptotic 
characteristic initial value problem} using formal series expansions in powers 
of $1/r$, where the coordinate $r$ is either the affine parameter of a future 
oriented null geodesic or a cleverly constructed
luminosity parameter. However, as was noted by some of the first researchers in
the area \cite{Fock}, combinations of powers of $\ln r$
and $1/r$ tend to arise naturally, even if we assume initially expansions in $1/r$.
 For long it was thought that these logarithmic 
terms were in some way connected to the 
presence of incoming radiation. In order to avoid the appearance of these 
logarithmic terms \emph{ad hoc}, some conditions were imposed on the 
\emph{Ansatz} for 
the asymptotic expansions. One of these conditions, used in the work by 
Bondi et al. \cite{vii}, \cite{ix} and Sachs \cite{viii} (Bondi-Sachs (BS) framework)
precluded the appearance of $1/r^2$ terms in some of the metric functions. 
This condition has
been known for a long time as the \emph{outgoing radiation condition for the
gravitational field }(ORC), because it bears some resemblance to
Sommerfeld's outgoing condition for the wave equation in flat space. It was later
realized that the ORC does not preclude the existence of incoming radiation. 
Moreover, the presence of incoming radiation in itself is not a problem. There
are some examples of spacetimes that contain a mixture of incoming and 
outgoing radiation and that have a somooth $\scri$. In any case, what 
has to be avoided (and is implicit in the definition of an isolated body) 
is the presence of radiation travelling ``infinitely long" 
distances (i.e. coming from past null infinity $\scri^-$). But this is 
achieved by imposing some boundary conditions directly on $\scri^-$ 
\cite{leipold}.

Newman and Penrose, working with their spin coefficients formalism (NP 
formalism), used a different condition. Namely, they assumed 
the $\Psi_0$ component of the Weyl tensor to behave asymptotically as 
$\O(r^{-5})$. This condition has its roots in the conformal compactification 
techniques introduced by Penrose. If one assumes $\Psi_0=\O(r^{-5})$ 
then one readily deduces the Peeling theorem, and all conformally rescaled 
quantities (the spin coefficients, the $\Psi$'s and the tetrad functions) are 
well defined at $\scri$. But if a slower decay for $\Psi_0$ is assumed
---say $\O(r^{-4})$--- then logarithmic terms will appear again in the 
asymptotic expansions. One also finds that some conformally rescaled 
quantities diverge at $\scri$. Hence one ends up with a null infinity 
that is non-smooth.

Not surprisingly, the ORC and Penrose's condition are found to be 
related. If the ORC does not holds (i.e. if we retain the ``nasty" $1/r^2$ 
terms in our initial expansions) then $\Psi_0=\O(r^{-4})$ \cite{javk98b}.

More recently, the work on the non-linear stability of Minkowski spacetime by 
Christodoulou and Klainerman has shown that generic spacetimes may exhibit a 
slower fall off (in some cases as low as $\O(r^{-7/2})$ on $\Psi_0$) and a non-smooth 
$\scri$. These results together with the fact that logarithmic terms 
have been observed by other authors in the context of cylindrically symmetric 
spacetimes \cite{ashtekar} and infinite rotating discs \cite{bicak} suggest that 
asymptotic expansions just in terms of powers of $1/r$ only are too restrictive for
many physically interesting applications. 

From the previous discussion, one can see that a natural way to generalize (and
hopefully gain more insight and better understanding!) the work that so far has 
been done on the asymptotic behaviour of the gravitational field is first, to 
assume a slower decay for the leading terms of the components of the Weyl tensor, 
and second, to allow the full appearance of logarithmic
terms in the asymptotic expansions\footnote{%
A precision should be done here. We will exclude from our study those
spacetimes that contain infinite series in powers of $\ln (r)$.}. Spacetimes 
that can be expanded asymptotically
in terms of a combination of powers of $1/r$ and $\ln r$ 
are known as \emph{polyhomogeneous spacetimes}.

An analysis of the solutions of the constraint equations by Andersson \& Chru\'sciel
\cite{andersson1}, \cite{andersson2}
seems also to indicate that an adequate framework to study the asymptotic 
gravitational field is that of the polyhomogeneous spacetimes. Chru\'sciel, MacCallum
and Singleton \cite{xiv} have shown that the hypothesis of 
polyhomogeneity is formally consistent with the Einstein field equations. 
Many results that hold in the non-polyhomogeneous setting can be recovered. The BMS group is
still the asymptotic symmetry
group, and  the uniqueness results for the Bondi mass of Chru\'sciel, Jezierski and
MacCallum hold \cite{chrusciel2}. However, the well posedness and existence/uniqueness of the 
asymptotic characteristic initial value problem is still an open question. Other 
classical results ---like the Peeling theorem---no longer hold, but an 
adequate generalization can be found in order to encompass the ``new" 
logarithmic terms \cite{xiv}.

Among this last group of results one has that of the Newman-Penrose constants,
a set of 10 absolutely conserved quantities that was found to exist for 
spacetimes satisfying the $\Psi_0=\O(r^{-5})$ condition. As shown in reference 
\cite{xiv} in a particular example, the Newman-Penrose constants are no 
longer constants for polyhomogeneous spacetimes. However, it was possible to 
show that for a restricted class of polyhomogeneous spacetimes (those 
spacetimes with a finite conformally rescaled shear $\widehat{\sigma}$ at 
$\scri$) it is possible to construct another set of 10 quantities (that 
we will call logarithmic Newman-Penrose constants) with a structure similar to 
that of the NP constants and that are conserved \cite{javk98a}. 

The objective of this article is to extend the work of references \cite{xiv} 
and \cite{javk98a} on these logarithmic constants, i.e. to find a family of 
functionals of $\Psi_0$ that for an arbitrary polyhomogeneous spacetime yields
conserved quantities, and for a non-polyhomogeneous spacetime reduces to the 
NP constants. In order to do so, it will be necessary to study in some 
detail the polyhomogeneous expansions of a spacetime. Some acquaintance with 
the NP spin coefficients formalism will be assumed in order to follow the 
calculations.

Much has been discussed about the interpretation and physical meaning of the 
NP constants: they have been regarded by some authors as the gravitational analogue of the 
electric charge, as they retain their values even after a process of 
radiation \cite{n-p65}, \cite{n-p68}.
However, the works trying to find an interpretation for them have been so far 
inconclusive, and some authors have even suggested on some grounds that the 
constants may be devoid of physical meaning \cite{bardeen}, \cite{press}, 
\cite{goldberg}. The new set of logarithmic 
constants adds to this old debate. Once their existence has been shown, the 
next natural step is to try to attach to them a physical meaning (if any!). In order to 
do a connection with previous work, stationary polyhomogeneous spacetimes are 
studied, and their logarithmic NP constants calculated. The result shows that 
for stationary spacetimes the logarithmic constants have a ``logarithmic 
quadrupole" structure.

This article is organized as follows: in section 2 we give a brief description
of the notations and conventions that are used in the present aerticle. In 
section 3 there is a discussion on how to calculate asymptotic expansion for 
polyhomogeneous spacetimes using the Newman-Penrose formalism. A description 
of the tetrad choice is also given. In section 4 we address the problem of how 
to generalize the Newman-Penrose constants to polyhomogeneous spacetimes 
(logarithmic NP constants). A detailed study of the evolution equation for the
$\Psi_0$ component of the Weyl tensor is done. Using this analysis, the main 
result of this article is proven: the existence of the so called logarithmic 
Newman-Penrose constants for all polyhomogeneous spacetimes. Finally, there is one appendix. In it one finds 2 tables that  summarize the behaviour of generic polyhomogeneous spacetimes. The results listed in these tables are used extensively in most of the arguments in this paper.

\section{Notation \& conventions.}

We will be working with spacetimes that formally can be expanded in terms of
powers of $1/r$ and $\ln r$. For example

\begin{equation}
\sigma =\sum_k\sigma _kr^{-k} =\sum_k \sum_j^{N_k} \sigma_{kj} r^{-k} \ln^{j
}r,
\end{equation}
where $\sigma _k$ denotes the polynomial in $z=\ln r$ with coefficients
depending on $u$, $\theta$ and $\varphi$ associated with $r^{-k}$. A second
subscript will mean the power of $z$ to which this function is attached. So, $%
\sigma _{i,j}$ is the coefficient that comes with $r^{-i}\ln ^jr$. For the
components of the Weyl tensor superscripts rather than
subscripts will be used. Let the symbol $\#$ denote the degree of a polynomial in $z$. We will
stick to the convention that the zero polynomial has degree $-\infty $. Hence

\begin{equation}
\#c(u,\theta,\varphi)=0,
\end{equation}

\begin{equation}
\#0=-\infty.
\end{equation}
A number in square brackets to the right of a polynomial will also indicate its
degree. For instance: $\sigma_2[N_3+1]$ means that $\#\sigma_2=N_3+1$. 

We will refer to the NP ``field equations'', the Bianchi identities and the
frame equations in the way they are listed in Stewart's book \cite{stewart}. In many
ocasions one will only be interested in the terms that go with a determinate
power of $1/r$. This will be indicated by attaching a subindex to the name of
the equation. A second subindex will refer to a particular power of $\ln r$.
So, for instance (Bb)$_{7,N_6}$ will mean that we are looking at the
relationship that goes with the $r^7\ln ^{N_6}$ term of the Bianchi identity
Bb.

When calculating the expansions of the diverse quantities, one usually needs 
to evaluate the derivatives and products of polyhomogeneous functions. The
following formulae are most useful in these circumstances:

\begin{equation} 
hg=\sum_{i=2}\sum_{k=1}^{i-1}h_k(z)g_{i-k}(z)r^{-i}, \label{multi}
\end{equation}

\begin{equation} 
\frac \partial {\partial r}h =\sum_{i=2}\left( h_{i-1}^{\prime 
}(z)-\left( i-1\right) h_{i-1}(z)\right) r^{-i}, \label{diff}
\end{equation}
where $h$ and $g$ are two arbitrary polyhomogeneous functions, the 
coefficients $h_k$, $g_k$ are polynomials in $z=\ln r$, and $^\prime$ denotes 
differentiation with respect to $z$.

Some use of the spin weighted spherical harmonics $(_s Y_{lm})$ and the $\eth$ 
and $\overline{\eth}$ (eth and ethbar)
operators will be made. In this case, we will also stick to Stewart's 
convention that is the same as the one in Penrose and Rindler
\cite{penroserindler}. Their connection with the NP directional derivatives is

\begin{eqnarray}
\eth &=& \delta + s(\overline{\alpha}-\beta), \\
\overline{\eth} &=& \overline{\delta} -s(\alpha -\overline{\beta}).
\end{eqnarray}

\section{Solving the NP equations using polyhomogeneous expansions.}

The first step into solving the NP field equations is to construct a coordinate
system and a null tetrad adapted to the problem. Suppose there is a one parameter family of null hypersurfaces. Denote this
family by $\N_u$. We can use the parameter $u$ to define a scalar field. Of
course, this parametrization is completely arbitrary, and one could have used
another parameter $\overline{u}$ as long as some smoothness
requirements are satisfied. On each of the geodesic generators $\gamma _u$
of the null hypersurfaces one can choose an affine parameter $r$. The null
hypersurfaces $\N_u$ intersect $\scri$ in cuts $\S_u$. On any one of the
cuts, say $\S_0$ we can choose arbitrary coordinates $x^\alpha $, $\alpha
=2,3 $. We can propagate the coordinates $x^\alpha $ on $\scri$ by
demanding $x^\alpha =const.$ on each generator of $\scri$. In this way we
have constructed a coordinate system $(u,r,x^\alpha )$ in the neighborhood
of $\scri$. There is some freedom left in this construction:

\begin{enumerate}
\item  we can always choose a different parameter $\overline{u}$ for the
family of null hypersurfaces;

\item  we can choose a different selection of coordinates $x^\alpha $ on $\S_0$, $%
\overline{x}^\alpha =\overline{x}^\alpha (x^\alpha )$;

\item  we can reset the origin and scaling of the affine parameter $r$, $%
\overline{r}=a(u,x^\alpha )r+b(u,x^\alpha )$.
\end{enumerate}

Once we have set our coordinate system, we can construct a null tetrad. An
obvious choice for the first element of the tetrad is the gradient of the
null hypersurfaces 

\begin{equation}
\mbox{l}=\mbox{d}u.
\end{equation}
The vector field $l^\mu$ is tangent to the geodesic generators $%
\gamma _u$ of $\N_u$. We can use the freedom of the scaling of the affine
parameter of these generators to set

\begin{equation}
l^\mu=\frac \partial {\partial r}.
\end{equation}
The freedom of the choice of the origin of the affine parameter will be used
so that $\rho $ will not contain any $1/r^2$ terms (see equation (\ref{rho2}%
)). Let us denote the 2-surfaces $u=const.$, $r=const$. by $\S_{u,r}$. It can be
seen that $l^\mu$ is future pointing and orthogonal to $\S_{u,r}$%
. There is only one other null direction with the same properties. The
second vector of our tetrad $n^\mu$ is chosen to be parallel to
this direction. The two other vectors of the tetrad, $m^\mu$
and $\overline{m}^\mu$, are constructed so that they span the
tangent space to $\S_{u,r}$, $\mbox{T}(\S_{u,r})$. So from this construction one deduces

\begin{equation}
n^\mu=\frac \partial {\partial u}+Q\frac \partial {\partial
r}+C^\alpha \frac \partial {\partial x^\alpha },
\end{equation}

\begin{equation}
m^\mu=\xi ^\alpha \frac \partial {\partial x^\alpha },
\end{equation}
so that the contravariant metric tensor is

\begin{equation} 
g_{NP}^{\mu \nu}=\left(  
\begin{array}{cccc} 
0 & 1 & 0 & 0 \\  
1 & 2Q & C^\theta & C^\varphi \\  
0 & C^\theta & -2 \xi^\theta \overline{\xi}^\theta & -\xi^\theta \overline{\xi}^
\varphi - \overline{\xi}^\theta \xi^\varphi \\
0 & C^\varphi & -\xi^\theta \overline{\xi}^
\varphi - \overline{\xi}^\theta \xi^\varphi & -2 \xi^\varphi 
\overline{\xi}^\varphi
\end{array} 
\right) , 
\end{equation} 
Using a spin-boost it is possible to set $\epsilon =\overline{\epsilon }$
everywhere. From the commutator equations we deduce

\begin{equation}
\kappa =\epsilon =0,
\end{equation}

\begin{equation}
\tau =\overline{\pi }=\overline{\alpha }+\beta \,
\end{equation}
and $\rho $ and $\mu $ are real.

The asymptotic characteristic initial value problem is usually set by 
supplying $\Psi_0$ on an initial null hypersurface $\N_0$, $\sigma$ on 
$\scri^+$, and $\Psi_1$, $\mbox{Re}\Psi_2$ and $\xi^i$ on $\Z = 
\scri^+ \cap \N_0$ \cite{kannar}.
 
The most general polyhomogeneous form for $\Psi_0$ that is physically 
reasonable to start with is 

\begin{equation}
\Psi_0 = \sum_{k=1} \Psi_0^{k},
\end{equation}
where the coefficients $\Psi_0^{k}$ are polynomials in $\ln r$, as mentioned 
before. With this initial data, the first two equations in the NP hierarchy to
be solved are
\begin{eqnarray}
D\rho = \rho^2 + \sigma\overline{\sigma}, \label{a} \\
D\sigma = 2\rho\sigma + \Psi_0.  \label{b}
\end{eqnarray}
Using (\ref{multi}) and (\ref{diff}) into (\ref{a}) and (\ref{b}) one obtains the 
following recurrence relations

\begin{equation} 
\rho _{i-1}^{\prime }-\left( i-1\right) \rho _{i-1}=\sum_{k=1}^{i-1}\left( 
\rho _k\rho _{i-k}+\sigma _k\overline{\sigma }_{i-k}\right) , 
\end{equation} 
\begin{equation} 
\sigma _{i-1}^{\prime }-\left( i-1\right) \sigma 
_{i-1}=2\sum_{k=1}^{i-1}\rho _k\sigma _{i-k}+\Psi _0^i.
\end{equation}

Setting $i=1$ we find that
\begin{equation} 
\Psi _0^1=0. 
\end{equation} 

For $i=2,$ the recurrence relations reduce to
\begin{eqnarray}
\rho^{\prime}_1 - \rho_1 = \rho_1^2 + \sigma_1 \overline{\sigma}_1, \label{a2} 
\\
\sigma^{\prime}_1 - \sigma_1 =2\rho_1\sigma_1 +\Psi_0^2. \label{b2}
\end{eqnarray}
From equation (\ref{a2}) we see that $\rho_1$ has to be a polynomial of degree 
zero. Couch \& Torrence \cite{couch} have shown that in order to have a 
spacetime with a Weyl tensor that vanishes at infinity then we require 
$\Psi_0=\O(r^{-2-\epsilon_1})$ where $\epsilon_1>0$. Hence, in order to have a 
physically realistic situation we set 
\begin{equation}
\Psi_0^2 =0,
\end{equation}
so that $\sigma_1$ is also of degree zero. The remaining system then 
reduces to
\begin{eqnarray}
- \rho_1 = \rho_1^2 + \sigma_1 \overline{\sigma}_1, \\
- \sigma_1 =2\rho_1\sigma_1.
\end{eqnarray}
which has 3 solutions: $\rho_1=0$, $\sigma_1=0$ (asymptotically plane) which 
yields degenerate spacetimes for which $m^\mu=\overline{m}^\mu=0$; $\rho_1= 
-\frac{1}{2}$ and $|\sigma_1|^2= \frac{1}{4}$ (asymptotically cylindrical)
which also produces degenerate spacetimes for which 
$m^\mu=\overline{m}^\mu$; and $\rho_1=-1$, $\sigma_1=0$ which gives rise to 
spacetimes that are asymptotically minkowskian. In what follows we will stick
to this last solution.

For $i=3$ one finds that 
\begin{eqnarray} 
\rho _2^{\prime } &=&0, \label{rho2} \\ 
\sigma _2^{\prime } &=&\Psi _0^3. \label{sigma2}
\end{eqnarray} 
Therefore $\rho _2$ is a polynomial of degree 0 in $\ln r$. Using the remaining 
freedom in the definition of the radial coordinate (recall that $r$ is an affine 
parameter) it is possible to 
redefine $r$ so that $\rho _2=0$ (see for example \cite{nu}). From 
equation (\ref{sigma2}) one reads
 
\begin{equation} 
\Psi _0^{3,j} =(j+1)\sigma _{2,j+1},  \label{shear1} 
\end{equation}
$j=0\ldots N_3$, 
whence $\#\sigma_2=N_3+1$. So, the only part of $\sigma_2$ that is not determined by the data on the 
initial hypersurface $\N$ is $\sigma_{2,0}$, in complete agreement 
with the results on the well posedness of the asymptotic characteristic initial 
value problem.
We can think of $\Psi _0^{3,j}$ $j=0...N-1$ as a contribution to the 
shear from the incoming radiation using Szekeres interpretation of the 
components of the Weyl tensor ($\Psi _0$ can be regarded as an incoming 
transverse wave)\cite{szekeres}. 
 
For $i=4$ one obtains,  
\begin{eqnarray}
\rho_3^{\prime}-\rho_3 = \sigma_2 \overline{\sigma}_2, \\
\sigma_3^{\prime}-\sigma_3= \Psi_0^{4}.
\end{eqnarray}
From these two equations one can readily see that $\#\rho_3=2N_3+2$ and 
$\#\sigma_3 = N_4$. Recurrence relations to calculate the coefficients of the 
two polynomials can be easily obtained. 

The process described here can be performed up to any desired order in $1/r$. A similar
analysis can be done with the remaining field equations, Bianchi identities 
and commutator relations. The order of solving the equations is given in 
reference \cite{stewart}.
A summary of this analysis is presented in the form 
of tables in Appendix A.

\section{Logarithmic NP constants for a generic polyhomogeneous spacetime.}

We are interested in the construction of conserved quantities that can be expressed 
as integrals over a cut of $\scri$, $S^2$. One expects these 
constants to be formed out of quantities that are initial data on the initial 
null hypersurface 
$\N_0$. As $\Psi_1^{4,0}$, $\mbox{Re}\Psi_2^{3,0}$, and $\sigma_{2,0}$ 
(the data that has to be prescribed on $\scri^+ \cap \N_0$) are 
used to construct the Bondi mass and the angular momentum of the spacetime
---quantities that are not conserved---, we do not expect to be 
able to construct conserved quantities out of them.

Newman \& Penrose \cite{n-p65}, \cite{n-p68}
have found that for non-polyhomogeneous spacetimes, 10 
conserved quantities can be constructed out of $\Psi_0$.
As it has been mentioned before, these quantities 
are not conserved for polyhomogeneous spacetimes.
Nevertheless, in the case when the leading term of the shear ($\sigma _2$)
contains no $\ln r$ terms (which corresponds to the situation $
\#\Psi_0^3= -\infty$) it  was possible to single out some other quantities with
a similar structure
(logarithmic NP constants) that are conserved. We want now to show
that even for the most general class of polyhomogeneous spacetimes it is
possible to construct such constants.

The key idea in the construction of these constants is to deduce from the Bianchi
identities an equation whose left hand side is a derivative with respect to 
the retarded time $u$
of one of the coefficients of $\Psi _0$, and whose right hand side is an $%
\overline{\eth }$ derivative of some combination of the coefficients in $%
\Psi _0,$ $\Psi _1,$ $\Psi _2$; that is, equations of the form

\begin{equation}
\dot{\Psi }_0^{i,M}=\overline{\eth }(F),  \label{np-equation}
\end{equation}
where \ $\dot{}$ \ denotes differentiation with respect to $u$, and the indices $i$ and $M$
are to be determined. Note that this 
equation has the same structure as the continuity equations in mechanics of 
continuous media: a derivative with respect to time plus the divergence of a 
flux, $F$ \footnote{The operator $\overline{\eth}\eth$ applied to a quantity of spin weight zero is equal to the angular part of the laplacian, $\nabla^2$. Hence, we can regard $\overline{\eth}\eth$ as the laplacian for spin weighted 
quantities, and $\eth$ or $\overline{\eth}$ as divergences (``square root of 
the laplacian'').}.
Equation (\ref{np-equation}) has  an overall spin weight 2, and the function $F$ spin weight 3.

We multiply equation (\ref{np-equation}) by
$(_2\overline{Y}_{lm})$, a spin 2 weighted spherical harmonic, in order to obtain an
equation of spin weight 0. Now, we integrate over the unit sphere (a cut of
$\scri$) so that

\begin{equation}
\frac \partial {\partial u}\left( \int_{S^2}\Psi _0^{i,M}(_2\overline{Y}%
_{lm})\mbox{d}S \right) =\int_{S^2}\overline{\eth }(F)(_2\overline{Y}%
_{lm})\mbox{d}S.
\end{equation}
Using the identity \cite{n-p65} 
\begin{equation}
\int_{S^2}\left( _s\overline{Y}_{l,m}\right) \overline{\eth }^{l-s+1}\zeta
\mbox{d}S =0,
\end{equation}
where $\zeta $ is of spin weight $l+1$ (in our case $\zeta =F$ has spin weight
3, hence $l=2$), we end up with 
 
\begin{equation}
\frac \partial {\partial u}\left( \int_{S^2}\Psi _0^{i,M}(_2\overline{Y}%
_{2m})\mbox{d}S \right) =0.
\end{equation}
Hence the 5 complex quantities

\begin{equation}
\mathcal{Q}^{i,M}=\int_{S^2}\Psi _0^{i,M}(_2\overline{Y}_{2m})\mbox{d}S ,
\end{equation}
$m=-2,-1 \ldots 2$ are conserved as long  as equation (\ref{np-equation}) holds.

The results in \cite{javk98a} show that one should expect to find the desired 
conservation laws in the coefficients of the highest power of $\ln r$ in the 
$1/r^6$ terms of the evolution equation for $\Psi_0$. An algebraic explanation
of this fact can be found by looking at the identity

\begin{equation}
\frac{\partial}{\partial r} \left( A(u,\theta, \varphi) r^{-i} \ln^{j} r 
\right) = (-i)A(u,\theta, \varphi)r^{-i-1}\ln^j r + jA(u,\theta, \varphi)r^{-i-1} 
\ln^{j-1} r,
\end{equation}
Hence, differentiation with respect to $r$ adds extra terms to the lower 
powers of $\ln r$, but not to the highest. Therefore, the coefficients of the 
highest power of $\ln r$  (for a given power of $1/r$) mimic the behaviour of 
the coefficients in a non-polyhomogeneous spacetime. The evolution equations 
in both cases will have the same algebraic form.

The situation discussed in this article is more elaborate than the one 
described in \cite{javk98a} as one has to deal with many new terms coming  
from 
$\Psi_0^3$. It will be necessary to keep track of the behaviour of the degrees
($N_i$) of the polynomials $\Psi_0^i$ as the system evolves. Therefore our 
discussion will start at the lowest order ($1/r^3$) instead of just going directly 
to the $1/r^6$ order terms.

As discussed previously, the most general form for $\Psi_0$ is the following:

\begin{equation}
\Psi _0=\Psi _0^3[N_3]r^{-3}+\Psi _0^4[N_4]r^{-4}+\Psi _0^5[N_5]r^{-5}+\Psi
_0^6[N_6]r^{-6}+\cdots,
\end{equation}
and its evolution equation is the Bianchi identity (Bb),

\begin{equation}
\Delta\Psi_0 -\delta\Psi_1 = 
(4\gamma-\mu)\Psi_0-2(2\tau+\beta)\Psi_1+3\sigma\Psi_2.
\end{equation}
One expects the values of $N_i$ to change as the system evolves. So, we can 
think of these numbers as discrete functions of the retarded time $u$ 
($N_i=N_i(u)$). A tilde \ $\widetilde{}$ \ over one of these numbers will mean 
that we are referring to its value at the initial null hypersurface $\N_0$.

The equation (Bb)$_3$ readily yields
\begin{equation}
\dot{\Psi}_0^3 =0,
\end{equation}
whence the polynomial $\Psi_0^3$ does not depend on $u$, and its degree 
($\#\Psi_0^3=\widetilde{N}_3=N_3$) is preserved during evolution.

From (Bb)$_4$ we obtain
\begin{equation}
\dot{\Psi}_0^4-\frac{1}{2}\Psi_0^{3
 \prime}[N_3-1]+\Psi_0^3[N_3]-\eth\Psi_1^3[N_3]=0,
\end{equation}
an equation of the form
\begin{equation}
\dot{\Psi}_0^4=P[N_3], \label{Bb4}
\end{equation}
$P[N_3]$ a polynomial of degree $N_3$.
If at the initial null hypersurface $\N_0$ we have $\widetilde{N}_4>N_3$ then we find 
that
\begin{equation}
\dot{\Psi}_0^{4,N_4}=\dot{\Psi}_0^{4,N_4-1}=\cdots=\dot{\Psi}_0^{4,N_3+1}=0,
\end{equation}
hence $\Psi_0^{4,N_4},\ldots,\Psi_0^{4,N_3+1}$ will be constants of motion, 
and the number $N_4$ a constant. The first evolution equation where the interaction
between $r^{-3}$ and $r^{-4}$ terms shows is
\begin{equation}
\dot{\Psi}_0^{4,N_3}+\Psi_0^{3,N_3}-\eth\Psi_1^{3,N_3}=0.
\end{equation}
Using, (Ba)$_4$
\begin{equation}
\Psi_1^{3\prime}[N_3-1] +\Psi_1^3[N_3]= \overline{\eth}\Psi_0^3[N_3],
\end{equation}
and the commutator for the $\eth$ derivatives we obtain
\begin{equation}
\dot{\Psi}_0^{4,N_3}+3 \Psi_0^{3,N_3}= \overline{\eth}\eth\Psi_0^{3,N_3} \label{psi04dot},
\end{equation}
so that we cannot derive a conservation law from this equation due to the
presence of the extra $3 \Psi_0^3$ term. If $N_3 \geq \widetilde{N}_4$ then 
the only thing we can affirm is that $N_3 \geq N_4$ (from (\ref{Bb4})).

The (Bb)$_5$ equation is much more involved. 
\begin{eqnarray}
\dot{\Psi}_0^5[N_5]-\frac{1}{2}(\Psi_0^{4\prime}-4\Psi_0^4) 
+Q_1[N_3+1](\Psi_0^{3\prime}-3\Psi_0^3)[N_3]+ 
C_2^\alpha[N_3+1]\partial_\alpha\Psi_0^3[N_3] \nonumber \\
-\delta_1\Psi_1^4[Y]+\delta_2[N_3+1]\Psi_1^3[N_3]= \nonumber \\
4\gamma_2[N_3+1]\Psi_0^3[N_3]+\frac{1}{2}\Psi_0^4[N_4]-\mu_2[N_3+1]\Psi_0^3[N_
3]-4\tau_2[N_3+1]\Psi_1^3[N_3] \nonumber \\
-2\beta_1\Psi_1^4[Y]-2\beta_2[N_3+1]\Psi_1^3[N_3
]+3\sigma_2[N_3+1]\Psi_2^3[N_3+1], \label{bb5}
\end{eqnarray}
an equation of the form
\begin{equation}
\dot{\Psi}_0^5=P[Y]
\end{equation}
where $Y=\max \{ N_4+1, 2N_3+2\}$. Note that if $\widetilde{N}_4+1>2N_3+2$ then
$\widetilde{N}_4>N_3$, and again due to (Bb)$_4$, $N_4$ will be a constant. 
Hence in this case $Y$ is a constant.
If in the other hand, $\widetilde{N}_4 +1 \leq 2N_3+2$ we end up with two cases
($ N_3<\widetilde{N}_4$ or $N_3 \geq \widetilde{N}_4$). If 
$N_3<\widetilde{N}_4$ then $N_4$ is a constant, and then the inequality $2N_3+
2 \geq N_4 +1$ holds for later times because it held at $\N_0$. 
And if $N_3 \geq \widetilde{N}_4$ then $N_3 \geq 
N_4$, and hence again $2N_3+2 \geq N_4 +1$. The conclusion is that $Y$ 
is a constant again.

Looking at equation (\ref{bb5}) we find that if $\widetilde{N}_5> \max\{N_4+1,2N_3+2\}$
we obtain a set of 
constants of motion ($\Psi_0^{5,N_5},\ldots,\Psi_0^{5,Y+1}$).
If $\widetilde{N_5} \leq Y=\max \{N_4+1,2N_3+2\}$ at $\N_0$ then we will not have 
constants of motion. The first equation showing an
interaction between terms of different 
orders in $r$ is equation (Bb)$_{5,Y}$:
\begin{equation}
\dot{\Psi}_0^{5,Y} =\eth\Psi_1^{4,Y} + \{\sigma_{2,N_3+1}\Psi_2^{3,N_3+1} \}, 
\label{bb5y}
\end{equation}
where the term in curly brackets $\{\}$ is present if $Y=2N_3+2$.
In order to obtain an equation in terms of ``initial data'' quantities only we 
require the equations (Ba)$_5$ and (Bc)$_4$:
\begin{equation}
\Psi_1^{4 \prime}[N_4]=\overline{\eth}\Psi_0^4[N_4] 
+(\overline{\delta}^2-4\alpha_2)[N_3+1]\Psi_0^3[N_3]+\pi_2[N_3+1]\Psi_0^3[N_
3]. \label{ba5}
\end{equation}
\begin{equation}
\Psi_2^{3 \prime}[N_3]-\overline{\eth}\Psi_1^3[N_3]=-\lambda_1[0]\Psi_0^3[N_3]
\label{bc4}
\end{equation}
Using equations (\ref{ba5}) and (\ref{bc4}) it is not hard to show that 
equation (\ref{bb5y}) does not have the required form.

Following \cite{n-p68} and \cite{javk98a} we expect to find conserved 
quantities of the desired form (if any) in the (Bb)$_6$ terms. The equation 
in this case is extremely involved, so it is better to look at it in parts:

\begin{eqnarray} 
\left[ \dot{\Psi }_0\right] _6 &=&\dot{\Psi }_0^6[N_6], \\ 
\left[ Q\partial_r\Psi _0\right] _6 &=&-\frac 12\left( \Psi 
_0^{5\prime }-5\Psi _0^5\right) [N_5]+Q_1[N_3+1]\left( \Psi _0^{4\prime 
}-4\Psi _0^4\right) [N_4] \nonumber \\
&&+Q_2[Y]\left( \Psi _0^{3\prime }-3\Psi _0^3\right) 
[N_3], \\ 
\left[ C^\alpha \partial _\alpha \Psi _0\right] _6 &=&C_2^\alpha 
[N_3+1]\partial _\alpha \Psi _0^4[N_4]+C_3^\alpha [Y]\partial _\alpha \Psi 
_0^3[N_3], \\ 
\left[ \delta \Psi _1\right] _6 &=&\delta _1\Psi _1^5[N_5]+\delta 
_2[N_3+1]\Psi _1^4[Y] \nonumber \\
&&+\delta _3[\max \{N_4,2N_3+2\}]\Psi _1^3[N_3], \\ 
\left[ \gamma \Psi _0\right] _6 &=&\gamma _2[N_3+1]\Psi _0^4[N_4]+\gamma 
_3[Y]\Psi _0^3[N_3], \\ 
\left[ \mu \Psi _0\right] _6 &=&-\frac 12\Psi _0^5[N_5]+\mu _2[N_3+1]\Psi 
_0^4[N_4]+\mu _3[Y]\Psi _0^3[N_3], \\ 
\left[ \tau \Psi _1\right] _6 &=&\tau _2[N_3+1]\Psi _1^4[Y]+\tau _3[Y]\Psi 
_1^3[N_3], \\ 
\left[ \beta \Psi _1\right] _6 &=&\beta _1\Psi _1^5[N_5]+\beta _2[N_3+1]\Psi 
_1^4[Y]+\beta _3[Y]\Psi _1^3[N_3], \\ 
\left[ \sigma \Psi _2\right] _6 &=&\sigma _2[N_3+1]\Psi _2^4[Y]+\sigma 
_3[N_4]\Psi _2^3[N_3+1]. 
\end{eqnarray}
Hence, (Bb)$_6$ is an equation of the form
\begin{equation}
\dot{\Psi}_0^6=P[X]
\end{equation}
where $X=\max\{N_5, 3N_3+3, N_3+N_4+2\}=\max\{N_5, (N_3+1)+Y\}$.
Now, if initially $\widetilde{N}_5 > (N_3+1)+Y$ then $N_5>Y$, and using 
(Bb)$_5$ we see that $N_5$ is a constant. On the other hand, if 
$\widetilde{N}_5 \leq (N_3+1)+Y$ then we end up with two possibilities: 
$\widetilde{N}_5>Y$ which yields $N_5$ constant; and $\widetilde{N}_5 \leq Y$ 
from which one deduces $N_5 \leq Y$. This arguments show the remarkable
fact that the number 

\begin{equation}
X=\max\{N_5, 3N_3+3, N_3+N_4+2\}
\end{equation}
is a constant independent of the initial values of $N_3$, $N_4$ and $N_5$.

As in previous orders, we will have a set
of constants of motion if $N_6>X=\max\{N_5, 3N_3+3, N_3+N_4+2\}$: 
$\Psi_0^{6,N_6},\ldots,\Psi_0^{6,X+1}$. The first non-trivial evolution 
equation will be

\begin{eqnarray}
\dot{\Psi }_0^{6,X}+\left\langle\frac 52\Psi _0^{5,X} \right\rangle +\left\langle\delta _1\Psi
_1^{5,X} \right\rangle +\left\{\delta^{2, N_3+1}\Psi _1^{4,Y}\right\}= \nonumber \\
\left\langle \frac{1}{2}\Psi _0^{5,X} \right\rangle -\left\{4\tau _{2,N_3+1}\Psi _1^{4,Y}
\right\}
-\left\langle 2\beta _1\Psi
_1^{5,X} \right\rangle -\left\{2\beta _{2,N_3+1}\Psi _1^{4,Y}\right\} \nonumber \\
+\left\{3\sigma _{2,N_3+1}\Psi _2^{4,Y}\right\} \label{Bb6}.
\end{eqnarray}
The terms in curly brackets $\{$ $\}$ will not be present if $X=N_5$, while
the terms in angle brackets $\langle$ $\rangle$ will not appear if $X=\max\{3N_3+3, 
N_3+N_4+2\}$.

In order to obtain an equation that contains only ``initial data'' quantities
we will  need to use the equations (Ba)$_6$ and (Bc)$_5$:

\begin{eqnarray}
\Psi _1^{5\prime }[N_5-1]-\Psi _1^5[N_5]-\overline{\eth }\Psi _0^5[N_5]-%
\overline{\delta }^2[N_3+1]\Psi _0^3[N_3]  \nonumber \\
-\overline{\delta }^3[\max
\{N_{4,}2N_3+2\}]\Psi _0^3[N_3] = \nonumber \\
\pi _2[N_3+1]\Psi _0^4[N_4]+\pi _3[Y]\Psi _0^3[N_3]-4\alpha _2[N_3+1]\Psi
_0^4[N_4] \nonumber \\
-4\alpha _3[Y]\Psi _0^3[N_3]+4\rho _3[2N_3+2]\Psi _0^3[N_3] \label{Ba6},
\end{eqnarray}
\begin{eqnarray}
\Psi _2^{4\prime }[Y-1]-\Psi _2^4[Y]-\overline{\eth }\Psi _1^4[Y]-%
\overline{\delta }^2[N_3+1]\Psi _1^3[N_3]=  \nonumber \\
-\lambda _1\Psi _0^4[N_4]-\lambda _2[N_3+1]\Psi _0^3[N_3]+2\pi
_2[N_3+1]\Psi _1^3[N_3]-2\alpha _2[N_3+1]\Psi _1^3[N_3] \label{Bc5}.
\end{eqnarray}
From equation (\ref{Ba6}) we obtain (when $X=N_5$)
\begin{equation}
\Psi_1^{5,N_5}=-\overline{\eth}\Psi_0^{5,N_5},
\end{equation}
and from (\ref{Bc5})
\begin{equation}
\Psi_2^{4,Y}=-\overline{\eth}\Psi_1^{4,Y}.
\end{equation}
Substitution of these results into equation (\ref{Bb6}) yields
\begin{eqnarray}
\dot{\Psi}_0^{6,X}&=& \left \langle \eth \overline{\eth}\Psi_0^{5,N_5} \right
\rangle -\left \langle 2\Psi_0^{5,N_5} \right \rangle - \left\{ 
(\delta^{2,N_3+1}-2\beta_{2,N_3+1})\Psi_1^{4,Y} \right\} \nonumber \\
 & &-\left\{ 4\tau_{2,N_3+1}\Psi_1^{4,y} \right \} 
-\left \{ 3\sigma_{2,N_3+1} \overline{\eth} 
\Psi_1^{4,Y} \right \}.
\end{eqnarray}
Using the techniques described in the previous section is not difficult (but 
very lenghty!) to obtain the values of the coefficients $\delta^{2,N_3+1}$, 
$\beta_{2,N_3+1}$, $\tau_{2,N_3+1}$, and  $\sigma_{2,N_3+1}$ in terms of 
$\Psi_0^3$:
\begin{eqnarray} 
\delta ^{2,N_3+1} &=&-\frac 1{N_3+1}\Psi _0^{3,N_3}\overline{\delta }^1, \\ 
\beta _{2,N_3+1} &=&\frac{-\alpha _1}{N_3+1}\Psi _0^{N_3}, \\ 
\tau _{2,N_3+1} &=&\frac 1{N_3+1}\overline{\eth }\Psi _0^{3,N_3}, \\ 
\sigma _{2,N_3+1} &=&\frac 1{N_3+1}\Psi _0^{3,N_3}. 
\end{eqnarray} 
With them and the commutator relation
\begin{equation}
(\overline{\eth}\eth-\eth\overline{\eth})\Psi_0^{5,X}=2\Psi_0^{5,X},
\end{equation}
one obtains an equation with the same structure as (\ref{np-equation}):
\begin{equation}
\dot{\Psi}_0^{6,X}=-\overline{\eth} \left( \left \langle\eth\Psi_0^{5,X} 
\right \rangle + 
\left \{\frac{4}{N_3+1}\Psi_0^{3,N_3}\Psi_1^{4,Y}\right \}  \right)
\label{lognp}.
\end{equation}
At this point, the relevance of the constancy of $X$ becomes clear. If it were 
not the case, then the continuity equation (\ref{lognp}) could change as the 
system evolves, and we would have different constants at different null 
hypersurfaces!

Therefore we have proven the following proposition:

\begin{proposition}
The quantities (5 complex in general) 
\begin{equation}
\mathcal{Q}_k^X=\int_{S^2}\Psi _0^{6,X}(_2\overline{Y}_{2k})\mbox{d}S \label{logconst}
\end{equation}
with $X=\max \{N_{5},3N_3+3,N_3+N_4+2\}$ are constants of motion for any
polyhomogeneous spacetime.
\end{proposition}

Observe that if $\widetilde{N}_3=\widetilde{N}_4=-\infty$ and $\widetilde{N}_5=0$
(that encompasses the 
non-polyhomogeneous case studied by Newman and Penrose) then $X=0$, and the 
logarithmic constants reduce to the NP constants. While if $N_3=-\infty$ then 
$X=N_5$ and we recover the logarithmic constants of \cite{javk98a}. 
Hence we have succeded in our
primary objective of finding a family of functionals of the quantity $\Psi_0$
that are absolutely conserved, and that reduce to the NP constants in the 
non-polyhomogeneous framework.

\section{Conclusions}
It has been shown that there is a functional of $\Psi_0$ that for a 
polyhomogeneous spacetime yields 10 absolutely conserved quantities, and for a
non-polyhomogeneous spacetime reduce to the NP constants. 

There are still some remaining open questions. One of them is related with the
well known fact that for linear fields one has an infinite hierarchy of NP 
constants. The general feeling in the case of the NP and logarithmic NP 
constants for the gravitational field is that the only quantities that one can
form that have the desired properties are precisely the ones already known. 
However a proof is still not available.

The other main question is that of the physical interpretation of the 
quantities. A step forward in this line would be the evaluation of the 
constants for some non-stationary spacetime. These topics will be the subject of 
future work.

\section*{Acknowlegments}
I would like to thank Prof M A H MacCallum for his continuous encouragement 
and advice. I also would like to thank Dr G D Kerr for numerous discussions on
spin weighted quantities and the $\eth$ and $\overline{\eth}$, 
and to Dr R Lazkoz for several suggestions. An important precision by Dr R Beig is also acknowledged. I am 
supported by the Consejo Nacional de Ciencia y Tecnolog\'{\i}a (CONACYT), Mexico, grant 110441/110491.

\appendix 

\section{Leading terms for the spin coefficients}
The objective of the present appendix is to present in a concise way the 
results of the calculations described in section 3. The tables give the 
degrees of the different polynomials that appear in the polyhomogeneous 
expansions. These tables are indispensable if one wishes to follow the 
calculations in full detail. As mentioned before, a $-\infty$ means that the 
given polynomial is zero.

\subsection{General polyhomogeneous spacetimes ($N_3\neq -\infty $).}

\[
\begin{tabular}{|c|c|c|c|c|c|}
\hline
& $1$ & $r^{-1}$ & $r^{-2}$ & $r^{-3}$ & $r^{-4}$ \\ \hline
$\Psi _0$ & $-\infty $ & $-\infty $ & $-\infty $ & $N_3$ & $N_4$ \\ \hline
$\Psi _1$ & $-\infty $ & $-\infty $ & $-\infty $ & $N_3$ & $\max
\{N_4+1,2N_3+2\}$ \\ \hline
$\Psi _2$ & $-\infty $ & $-\infty $ & $-\infty $ & $N_3+1$ & $\max
\{N_4+1,2N_3+2\}$ \\ \hline
$\Psi _3$ & $-\infty $ & $-\infty $ & $0$ &  &  \\ \hline
$\Psi _4$ & $-\infty $ & $0$ & $0$ &  &  \\ \hline
$\rho $ & $-\infty $ & $0$ & $-\infty $ & $2N_3+2$ & $N_3+N_4+1$ \\ \hline
$\sigma $ & $-\infty $ & $-\infty $ & $N_3+1$ & $N_4$ & \\ \hline
$\alpha $ & $-\infty $ & $0$ & $N_3+1$ & $\max \{N_4+1,2N_3+2\}$ &  \\ \hline
$\beta $ & $-\infty $ & $0$ & $N_3+1$ & $\max \{N_4+1,2N_3+2\}$ &  \\ \hline
$\tau $ & $-\infty $ & $-\infty $ & $N_3+1$ & $\max \{N_4+1,2N_3+2\}$ &  \\ 
\hline
$\pi $ & $-\infty $ & $-\infty $ & $N_3+1$ & $\max \{N_4+1,2N_3+2\}$ &  \\ 
\hline
$\gamma $ & $-\infty $ & $-\infty $ & $N_3+1$ & $\max \{N_4+1,2N_3+2\}$ & 
\\ \hline
$\lambda $ & $-\infty $ & $0$ & $N_3+1$ & $\max \{N_4+1,2N_3+2\}$ &  \\ 
\hline
$\mu $ & $-\infty $ & $0$ & $N_3+1$ & $\max \{N_4+1,2N_3+2\}$ &  \\ \hline
$\nu $ & $-\infty $ & $-\infty $ & $N_3+1$ & $\max \{N_4+1,2N_3+2\}$ &  \\ 
\hline
$\xi ^\alpha $ & $-\infty $ & $0$ & $N_3+1$ & $\max \{N_4,2N_3+2\}$ &  \\ 
\hline
$C^\alpha $ & $-\infty $ & $-\infty $ & $N_3+1$ & $\max \{N_4+1,2N_3+2\}$ & 
\\ \hline
$Q$ & $0$ & $N_3+1$ & $\max \{N_4+1,2N_3+2\}$ &  &  \\ \hline
\end{tabular}
\]

\subsection{Polyhomogeneous space-times with a finite shear at $\scri$.}

\[
\begin{tabular}{|c|c|c|c|c|c|c|c|}
\hline
& $1$ & $r^{-1}$ & $r^{-2}$ & $r^{-3}$ & $r^{-4}$ & $r^{-5}$ & $r^{-6}$ \\ 
\hline
$\Psi _0$ & $-\infty $ & $-\infty $ & $-\infty $ & $-\infty $ & $N_4$ & $N_5$
& $N_6$ \\ \hline
$\Psi _1$ & $-\infty $ & $-\infty $ & $-\infty $ & $-\infty $ & $N_4+1$ & $%
N_5$ &  \\ \hline
$\Psi _2$ & $-\infty $ & $-\infty $ & $-\infty $ & $0$ & $N_4+1$ & $N_5$ & 
\\ \hline
$\Psi _3$ & $-\infty $ & $-\infty $ & $0$ & $0$ &  &  &  \\ \hline
$\Psi _4$ & $-\infty $ & $0$ & $0$ &  &  &  &  \\ \hline
$\rho $ & $-\infty $ & $0$ & $-\infty $ & $0$ & $N_4$ &  &  \\ \hline
$\sigma $ & $-\infty $ & $-\infty $ & $0$ & $N_4$ & $N_5$ &  &  \\ \hline
$\alpha $ & $-\infty $ & $0$ & $0$ & $N_4+1$ &  &  &  \\ \hline
$\beta $ & $-\infty $ & $0$ & $0$ & $N_4+1$ &  &  &  \\ \hline
$\tau $ & $-\infty $ & $-\infty $ & $0$ & $N_4+1$ &  &  &  \\ \hline
$\pi $ & $-\infty $ & $-\infty $ & $0$ & $N_4+1$ &  &  &  \\ \hline
$\gamma $ & $-\infty $ & $-\infty $ & $0$ & $N_4+1$ &  &  &  \\ \hline
$\lambda $ & $-\infty $ & $0$ & $0$ & $N_4+1$ &  &  &  \\ \hline
$\mu $ & $-\infty $ & $0$ & $0$ & $N_4+1$ &  &  &  \\ \hline
$\nu $ & $-\infty $ & $-\infty $ & $0$ & $N_4+1$ &  &  &  \\ \hline
$\xi ^\alpha $ & $-\infty $ & $0$ & $0$ & $N_4$ &  &  &  \\ \hline
$C^\alpha $ & $-\infty $ & $-\infty $ & $0$ & $N_4+1$ &  &  &  \\ \hline
$Q$ & $0$ & $0$ & $N_4+1$  &  &  &  &  \\ \hline
\end{tabular}
\]

\end{document}